\documentclass[12pt,a4paper]{article}
\pdfoutput=1
\usepackage{color}
\usepackage{amssymb,amsmath,bm,bbm}
\usepackage{epsf}
\usepackage{epsfig}
\usepackage{afterpage}
\usepackage{longtable}
\usepackage[dvipsnames]{xcolor}
\usepackage[linktoc=page,bookmarks=false,colorlinks=false,
linkbordercolor=RoyalBlue,citebordercolor=ForestGreen,urlbordercolor=CornflowerBlue]{hyperref}
\usepackage{latexsym,mathrsfs,dsfont}
\usepackage[normalem]{ulem} % for strikeout \sout
\usepackage[compress]{cite}
\usepackage{graphicx}
\usepackage{url}
\usepackage{paralist}
\usepackage{bbold}

\newcommand{\ord}{\mathcal{O}}

\newcommand{\IM}{{\rm Im}}
\newcommand{\RE}{{\rm Re}}

\newcommand{\gev}{\, {\rm GeV}}
\newcommand{\mev}{\, {\rm MeV}}

\newcommand{\vcb}{|V_{cb}|}

\newcommand{\vub}{|V_{ub}|}

\newcommand{\bsi}{B_6^{(1/2)}}
\newcommand{\bei}{B_8^{(3/2)}}

\def\epe{\varepsilon'/\varepsilon}
\newcommand{\beq}{\begin{equation}}
\newcommand{\eeq}{\end{equation}}
\newcommand{\be}{\begin{equation}}
\newcommand{\ee}{\end{equation}}
\newcommand{\bi}{\begin{itemize}}
\newcommand{\ei}{\end{itemize}}
\newcommand{\ba}{\begin{array}}
\newcommand{\ea}{\end{array}}
\newcommand{\beqa}{\begin{eqnarray}}
\newcommand{\eeqa}{\end{eqnarray}}
\newcommand{\bea}{\begin{eqnarray}}
\newcommand{\eea}{\end{eqnarray}}
\newcommand{\beqn}{\begin{eqnarray}}
\newcommand{\eeqn}{\end{eqnarray}}

\definecolor{red}{cmyk}{0,1,1,0.4}

\usepackage{fancyhdr}
\advance \headheight by 15.0truept       % for 12pt mandatory...
\begin{document}

%\begin{flushleft}
%{\em Version of 20 July 2015}
%\end{flushleft}

\vspace{-14mm}
\begin{flushright}
        {FLAVOUR(267104)-ERC-101}\\
CP3-15-19
\end{flushright}

\vspace{8mm}

\begin{center}
{\Large\bf
\boldmath{Upper Bounds on $\epe$ Parameters $\bsi$ and $\bei$ \\ from  
 Large $N$ QCD and other  News}}
\\[12mm]
{\bf \large  Andrzej~J.~Buras ${}^a$ and Jean-Marc G\'erard${}^b$ \\[0.8cm]}
{\small
${}^a$TUM Institute for Advanced Study, Lichtenbergstr.~2a, D-85748 Garching, Germany\\
Physik Department, TU M\"unchen, James-Franck-Stra{\ss}e, D-85748 Garching, Germany\\[2mm]
${}^b$ Centre for Cosmology,
Particle Physics and Phenomenology (CP3), Universit{\'e} catholique de Louvain,
Chemin du Cyclotron 2,
B-1348 Louvain-la-Neuve, Belgium}
\end{center}

\vspace{8mm}

\abstract{%
\noindent
We demonstrate that in the large $N$ approach  developed by the authors in collaboration with Bardeen, the parameters $\bsi$ and $\bei$ parametrizing the $K\to\pi\pi$ matrix elements $\langle Q_6 \rangle_0$ and $\langle Q_8 \rangle_2$ of the dominant QCD and electroweak operators  receive both {\it negative} $\ord(1/N)$ corrections such that $\bsi \le \bei<1 $ in agreement with the recent lattice results of the RBC-UKQCD collaboration. 
We also point out that the pattern of the size of  the hadronic matrix elements of all QCD and electroweak penguin operators $Q_i$ contributing to the $K\to \pi \pi$ amplitudes $A_0$ and $A_2$, obtained  by this  lattice collaboration, provides further  support to our large $N$ approach. In particular, the lattice result for the matrix element 
$\langle Q_8 \rangle_0$ implies for the corresponding parameter $B_8^{(1/2)}=1.0\pm 0.2$ to be compared with large $N$ value $B_8^{(1/2)}=1.1 \pm 0.1$. 
We discuss briefly the implications of these findings for the ratio 
$\epe$. In fact, with the precise value for $\bei$ from 
 RBC-UKQCD collaboration, our upper bound on $\bsi$ implies $\epe$ in the SM roughly by a factor of two below its
experimental value $(16.6\pm 2.3)\times 10^{-4}$. We also briefly 
comment on the parameter $\hat B_K$ and the $\Delta I=1/2$ rule. }

\setcounter{page}{0}
\thispagestyle{empty}
\newpage

\tableofcontents

\section{Introduction}
The decays $K\to \pi\pi$  have played a very important 
role since their discovery in the 1950s, both in the construction of the Standard 
Model (SM) and more recently in the tests of its possible extensions. Most 
of the discussions in the literature centred on the following quantities:
\begin{itemize}
\item
The ratio 
\be\label{N1a}
\frac{{\rm Re}A_0}{{\rm Re}A_2}=22.4\,,
\ee
which expresses the so-called $\Delta I=1/2$ rule \cite{GellMann:1955jx,GellMann:1957wh}.
\item
The parameter $\varepsilon_K$, a measure of indirect CP-violation in 
$K_L\to\pi\pi$ decays, found to be
\be\label{N2}
\varepsilon_K=2.228(11)\times 10^{-3}e^{i\phi_\varepsilon},
\ee
where $\phi_\varepsilon=43.51(5)^\circ$. 
\item
The ratio of the direct CP-violation and indirect CP-violation in $K_L\to\pi\pi$ decays  measured 
to be \cite{Beringer:1900zz,Batley:2002gn,AlaviHarati:2002ye,Worcester:2009qt}
\be\label{eprime}
\RE(\epe)=(16.6\pm 2.3)\times 10^{-4}.
\ee
\end{itemize}

 Unfortunately, due to non-perturbative uncertainties originating in the 
hadronic matrix elements of contributing four-quark operators, it took a long 
time to obtain meaningful results for all these observables in QCD.
But already in the second half of the 1980s, we have developed an approach 
to $K^0-\bar K^0$ mixing and non-leptonic $K$-meson decays 
\cite{Buras:1985yx,Bardeen:1986vp,Bardeen:1986uz,Bardeen:1986vz,Bardeen:1987vg}  based on the dual representation of QCD as a theory of 
weakly interacting mesons for large $N$, where $N$ is the number of colours 
 \cite{'tHooft:1973jz,'tHooft:1974hx,Witten:1979kh,Treiman:1986ep}.
 The most recent results from our approach can be found in  \cite{Buras:2014maa,Buras:2014apa}.

This approach provided, in particular, first results within QCD for the 
 amplitudes ${\rm Re}A_0$ and ${\rm Re}A_2$ 
in the ballpark of experimental values \cite{Bardeen:1986vz}. In this manner, 
for the first time, the SM dynamics behind the  $\Delta I=1/2$ rule 
has been identified. In particular, it has been emphasized that at 
scales $\ord(1\gev)$ long distance dynamics in hadronic matrix elements 
of current-current operators and not QCD-penguin operators, as originally proposed in 
\cite{Shifman:1975tn}, are dominantly 
responsible for 
this rule. Moreover, it has been demonstrated analytically why ${\rm Re}A_0$ is 
enhanced and why ${\rm Re}A_2$ is suppressed relative to the vacuum insertion  approximation (VIA) estimates. In this context, we have emphasized that the so-called Fierz terms in the latter approach totally misrepresent $1/N$ corrections to the strict large $N$ limit for these amplitudes \cite{Bardeen:1986vz}.

Our approach, among other applications, allowed us to consistently calculate, for the first time within QCD, the non-perturbative parameters $\hat B_K$, $\bsi$ and 
$\bei$ 
governing the corresponding matrix elements of $\Delta S=2$ SM current-current operator and $K\to\pi\pi$ matrix elements of the 
dominant QCD-penguin ($Q_6$) and  electroweak penguin ($Q_8$)
operators. These parameters are crucial for the evaluation of $\varepsilon_K$ and $\epe$ within the SM and its various 
extensions. 
Other applications of large $N$ ideas 
to $K\to\pi\pi$  and $\hat B_K$, but in a different spirit than our original approach, are reviewed in \cite{Cirigliano:2011ny}.  We will comment in Section~\ref{otherN} on those which reached very different conclusions from ours.

It is interesting and encouraging that most of our results have been confirmed 
by several recent lattice QCD calculations that we will specify below. While the lattice QCD approach has a better control over the errors than our approach, it does 
not provide the physical picture of the dynamics behind the obtained numerical 
results. This is in particular seen in the case of the $\Delta I=1/2$ rule 
where our  analytic approach  offers a very simple picture of 
 the dynamics behind this rule, as summarized again in  \cite{Buras:2014maa,Buras:2014apa}.

In the present paper, we briefly compare in Section~\ref{sec:1} the status of lattice results for 
$\hat B_K$ and the $\Delta I=1/2$ rule with the ones obtained in our 
approach. Subsequently, in Section~\ref{sec:2}  we demonstrate that the pattern of 
the size of the matrix elements for penguin operators  presented recently 
by the RBC-UKQCD collaboration  for $A_0$  \cite{Bai:2015nea} and $A_2$ amplitudes \cite{Blum:2015ywa} gives another support 
 to our approach. In Section~\ref{sec:3}, we derive {upper} bounds on the parameters 
$\bsi$ and $\bei$ and discuss {briefly} in Section~\ref{sec:5} their phenomenological implications for $\epe$. In Section~\ref{otherN} we describe briefly  the results obtained in other large $N$ QCD approaches. An outlook 
is presented in Section~\ref{sec:4}.

\boldmath
\section{$\hat B_K$ and the $\Delta I=1/2$ Rule}\label{sec:1}
\unboldmath
\boldmath
\subsection{$\hat B_K$}
\unboldmath
The scale and renormalization scheme dependent parameter $B_K(\mu)$ is related 
to the relevant hadronic matrix element of the $\Delta S=2$ operator
\be\label{DS2OP}
Q=(\bar sd)_{V-A}(\bar s d)_{V-A}
\ee
as follows\footnote{In this paper, we use the normalization of weak decay constants given in (\ref{FpFK}).}
\be\label{eq:5}
\langle \bar K^0|Q(\mu)|K^0\rangle = B_K(\mu)\frac{8}{3}F_K^2 m_K^2.
\ee
More useful is the renormalization group invariant 
parameter $\hat B_K$ that is given by \cite{Buras:1990fn}
\be\label{BKhat}
\hat B_K=B_K(\mu)\left[\alpha_s^{(3)}(\mu)\right]^{-d}
\left[1+\frac{\alpha_s^{(3)}(\mu)}{4\pi}J_3\right], \qquad d=\frac{9(N-1)}{N(11N-6)}\,.
\ee
We have shown the $N$-dependence of the exponent $d$ in the leading term to 
signal that $d$ vanishes in the large $N$ limit. The coefficient $J_3$ is renormalization scheme dependent. This dependence cancels the one of $B_K(\mu)$.

 As in the strict 
large $N$ limit the exponent in (\ref{BKhat}) and the NLO term involving 
$J_3$ vanish, one finds  \cite{Buras:1985yx} that 
independently of any renormalization scale or renormalization scheme for 
the operator $Q$ 
\be\label{BKLO}
\hat B_K \rightarrow 0.75, \qquad ({\rm in~large~N~limit,~1986}).
\ee

It can be shown that including $1/N$ corrections suppresses $\hat B_K$ so
that  \cite{Gerard:2010jt} 
\be\label{Bound}
\hat B_K\le 0.75~,\qquad  ({\rm in~1/N~ expansion}).
\ee
Our latest analysis in {our approach} gave   \cite{Buras:2014maa}
\be\label{BKfinal}
\hat B_K= 0.73\pm 0.02, \qquad  ({\rm in~dual~ QCD}),
\ee
where the error should not be considered as a standard deviation. Rather, 
this result represents the range for $\hat B_K$ we expect in our approach after 
the inclusion of NLO QCD corrections and the contributions of pseudoscalar and 
vector mesons as discussed in detail in \cite{Buras:2014maa}.

On the other hand, the world lattice average for 
$\hat B_K$ based on the calculations of various groups \cite{Aoki:2010pe,Bae:2010ki,Constantinou:2010qv,Colangelo:2010et,Bailey:2012bh,Durr:2011ap} 
reads for $N_f=2+1$ calculations (recent FLAG update of \cite{Colangelo:2010et})
\be\label{L2012}
\hat B_K=0.766\pm 0.010, \qquad   ({\rm in~ lattice~QCD,}~2014).
\ee
See also the recent analyses in \cite{Frison:2013fga,Bae:2014sja}.
While this result violates the bound in (\ref{Bound}), it should be noted that
a number of lattice groups among \cite{Aoki:2010pe,Bae:2010ki,Constantinou:2010qv,Colangelo:2010et,Bailey:2012bh,Durr:2011ap} published results with central 
values satisfying the bound in (\ref{Bound}) but the errors did not allow 
for a clear cut conclusion. 
In fact, the most recent update from staggered 
quarks \cite{Bae:2014sja} quotes precisely $\hat B_K=0.738\pm0.005$ but additional systematic error of $0.037$ does not allow for definite conclusions. Similarly, the Rome group \cite{Carrasco:2015pra} finds basically the result in (\ref{BKfinal}).
 We expect therefore that improved lattice calculations 
will satisfy our  bound one day and in a few years from now lattice average for 
$\hat B_K$ will read $\hat B_K\approx 0.74$.

Finally, let us remark that while the lattice approach did not provide the 
explanation why $\hat B_K$ is so close to its large $N$ limit $0.75$, in our 
approach the smallness of $1/N$ corrections follows from the approximate 
cancellation of negative pseudoscalar meson contributions by the positive 
vector meson contributions.

\boldmath
\subsection{$\Delta I=1/2$ Rule}
\unboldmath
A very detailed comparison of the calculations of ${\rm Re}A_0$ and 
${\rm Re}A_2$ in our approach and the lattice QCD has been presented in 
 \cite{Buras:2014maa} and our present discussion is meant to be an update 
due to new results of the RBC-UKQCD collaboration on  ${\rm Re}A_0$  \cite{Bai:2015nea}.

First, let us mention that both the dual approach to QCD and lattice approach 
obtain satisfactory results for the amplitude  ${\rm Re}A_2$. On the other 
hand, whereas we find  \cite{Buras:2014maa}
\be\label{DRULEN}
\left(\frac{{\rm Re}A_0}{{\rm Re}A_2}\right)_{{\rm dual~QCD}}=16.0\pm1.5 \,,
\ee
the most recent result from the RBC-UKQCD collaboration reads \cite{Bai:2015nea}
\be\label{DRULEL}
\left(\frac{{\rm Re}A_0}{{\rm Re}A_2}\right)_{{\rm lattice~QCD}}=31.0\pm6.6 \,.
\ee

Due to large error in the lattice result, both results are compatible with 
each other and both signal that this rule follows dominantly from the 
QCD dynamics related to current-current operators. But our approach, being 
analytic, allows to connect the $\Delta I=1/2$ rule to the main properties of QCD: 
asymptotic freedom and the related evolutions of weak matrix elements 
which at long distance scales can be performed in the dual representation 
of QCD as a theory of weakly interacting mesons for large $N$.
As lattice QCD calculations are performed  basically at a single energy scale, no such physical explanation of this rule is expected from that framework. 
To this end, lattice calculations would have to be performed at scales below 
$1\gev$ which is straightforward in our approach but appears impossible by 
lattice methods at present.

On the other hand,  from the present perspective 
only lattice simulations 
can provide  precise value of  ${\rm Re}A_{0}$ one day, so that we will know 
whether some part of this rule at the level of $(20-30)\%$, as signalled 
by the result in (\ref{DRULEN}), originates in new physics (NP) contributions. Indeed, as demonstrated in \cite{Buras:2014sba},  a heavy $Z^\prime$ and in particular a heavy $G^\prime$ in the reach of the 
LHC could be responsible for the missing piece in ${\rm Re}A_0$ in 
(\ref{DRULEN}). On the basis of the analysis in \cite{Buras:2014sba} { it is much harder to  
 bring this ratio with the help of NP from 31 down to 22 without 
violating $\Delta M_K$ constraint,  but this requires a separate study. }

Of some interest is the ratio of the matrix elements $\langle Q_2 \rangle_0$
and $\langle Q_1 \rangle_0$.  It equals $-2$ in the large $N$ limit, 
corresponding to $\mu=0$ \cite{Buras:2014maa}. Evolving these matrix elements 
to $\mu= 1\gev$ in the meson theory and subsequently to $\mu=1.53\gev$ in 
the quark theory, we 
find in the NDR-${\rm \overline{MS}}$ scheme\footnote{We thank Martin Gorbahn for checking this result.}
\be
\frac{\langle Q_2 \rangle_0}{\langle Q_1 \rangle_0}=-1.50\pm 0.10,\qquad \mu=1.53\gev,\qquad {\rm (dual~QCD)}.
\ee
The corresponding result in \cite{Bai:2015nea} reads
\be
\frac{\langle Q_2 \rangle_0}{\langle Q_1 \rangle_0}=-1.12\pm 0.49,\qquad \mu=1.53\gev,\qquad {\rm (lattice~QCD)}.
\ee

In view of large uncertainty in the lattice result, these two ratios are compatible with each other. We expect on the basis of the results in (\ref{DRULEN}) and
(\ref{DRULEL}) that this ratio will be eventually found in the ballpark of 
$-1.4$.

\boldmath
\section{Matrix Elements of Penguin Operators}\label{sec:2}
\unboldmath

\subsection{Preliminaries}
We will consider the usual basis of operators contributing to $K\to \pi\pi$ 
amplitudes \cite{Buras:1993dy}, namely

{\bf Current--Current:}
\begin{equation}\label{O1} 
Q_1 = (\bar s_{\alpha} u_{\beta})_{V-A}\;(\bar u_{\beta} d_{\alpha})_{V-A}\,,
~~~~~~Q_2 = (\bar su)_{V-A}\;(\bar ud)_{V-A} 
\end{equation}

{\bf QCD--Penguins:}
\begin{equation}\label{O2}
Q_3 = (\bar s d)_{V-A}\sum_{q=u,d,s}(\bar qq)_{V-A},~~~~~~   
 Q_4 = (\bar s_{\alpha} d_{\beta})_{V-A}\sum_{q=u,d,s}(\bar q_{\beta} 
       q_{\alpha})_{V-A} 
\end{equation}
\begin{equation}\label{O3}
 Q_5 = (\bar s d)_{V-A} \sum_{q=u,d,s}(\bar qq)_{V+A},~~~~~  
 Q_6 = (\bar s_{\alpha} d_{\beta})_{V-A}\sum_{q=u,d,s}
       (\bar q_{\beta} q_{\alpha})_{V+A} 
\end{equation}

{\bf Electroweak Penguins:}
\begin{equation}\label{O4} 
Q_7 = \frac{3}{2}\;(\bar s d)_{V-A}\sum_{q=u,d,s}e_q\;(\bar qq)_{V+A}, 
~~~~~ Q_8 = \frac{3}{2}\;(\bar s_{\alpha} d_{\beta})_{V-A}\sum_{q=u,d,s}e_q
        (\bar q_{\beta} q_{\alpha})_{V+A}
\end{equation}
\begin{equation}\label{O5} 
 Q_9 = \frac{3}{2}\;(\bar s d)_{V-A}\sum_{q=u,d,s}e_q(\bar q q)_{V-A},
~~~~Q_{10} =\frac{3}{2}\;
(\bar s_{\alpha} d_{\beta})_{V-A}\sum_{q=u,d,s}e_q\;
       (\bar q_{\beta}q_{\alpha})_{V-A} 
\end{equation}
Here, $\alpha,\beta$ denote colour indices and $e_q$ denotes the electric quark
charges reflecting the electroweak origin of $Q_7,\ldots,Q_{10}$. Finally,
$(\bar sd)_{V-A}\equiv \bar s_\alpha\gamma_\mu(1-\gamma_5) d_\alpha$ as in (\ref{DS2OP}).

Recently, the RBC-UKQCD collaboration published their results for the 
matrix elements $\langle Q_i\rangle_0$  \cite{Bai:2015nea}. 
Their matrix elements are given for {\it three} dynamical quarks at $\mu=1.53\gev$, which is too high for the 
direct comparison with our approach in the case of current-current operators.
 On the other hand, the parameters $\bsi$ and  $\bei$ of 
the QCD penguin operator $Q_6$ and the electroweak penguin operator $Q_8$ 
are known \cite{Buras:1993dy} to be practically scale independent for $1.0\gev\le \mu\le 3.0\gev$. Therefore these 
results constitute a useful test of our approach. Another issue is the 
colour suppression of some matrix elements of other penguin operators which 
is predicted within our approach. 
We would like to check whether the pattern of this suppression is also seen 
in the lattice data. 

\subsection{Hadronic matrix elements}\label{sec:3.2}
The hadronic matrix elements of operators $Q_i$ that are most useful for our
discussions are 
\be
\langle Q_i \rangle_I \equiv
\langle \left(\pi\pi\right)_I \left| Q_i \right| K \rangle \, ,
\label{eq:5.1}
\ee
with $I=0,2$ being strong isospin. 

It should be recalled that for $\mu\le m_c$, when charm quark has been integrated out, only  seven of the operators listed above are independent of each 
other. Eliminating then $Q_4$, $Q_9$ and $Q_{10}$ in terms of the remaining 
seven operators allows, in the isospin symmetry limit, to find the following 
 important relations  \cite{Buras:1993dy}\footnote{In writing (\ref{Q4}) we neglect a small $\ord(\alpha_s)$ correction in the NDR scheme which is explicitly 
given in (4.44) of \cite{Buras:1993dy}.}
\begin{eqnarray}
\langle Q_4 \rangle_0 &=&  \langle Q_3 \rangle_0 + \langle Q_2 \rangle_0
                          -\langle Q_1 \rangle_0 \,, \label{Q4}\\
\langle Q_9 \rangle_0 &=&
\frac{3}{2} \langle Q_1 \rangle_0 - \frac{1}{2} \langle Q_3 \rangle_0 \, ,
\label{eq:5.12} \\
\langle Q_{10} \rangle_0 &=&
    \langle Q_2 \rangle_0 + \frac{1}{2} \langle Q_1 \rangle_0
  - \frac{1}{2} \langle Q_3 \rangle_0 \, ,
\label{eq:5.13}\\
\langle Q_9 \rangle_2 &=&
   \langle Q_{10} \rangle_2 = \frac{3}{2} \langle Q_1 \rangle_2 \, ,
\label{eq:5.18}
\end{eqnarray}
where we have used
\begin{eqnarray}
\langle Q_1 \rangle_2 &=&
\langle Q_2 \rangle_2  \, .
\label{eq:5.14} 
\end{eqnarray}
We have checked that these relations have been used in  \cite{Bai:2015nea}.

Of particular importance for our discussion are the matrix elements
\begin{eqnarray}\label{eq:Q60}
\langle Q_6(\mu) \rangle_0 &=&-\, h
\left[ \frac{2 m_{\rm K}^2}{m_s(\mu) + m_d(\mu)}\right]^2 (F_K-F_\pi)
\,B_6^{(1/2)}\,,\\
\label{eq:Q82}
\langle Q_8(\mu) \rangle_2 &=&\frac{h}{2\sqrt{2}}
\left[ \frac{2 m_{\rm K}^2}{m_s(\mu) + m_d(\mu)}\right]^2 F_\pi \,B_8^{(3/2)}\,,
\\
\label{eq:Q80}
\langle Q_8(\mu) \rangle_0 &=& \frac{h}{2}
\left[ \frac{2 m_{\rm K}^2}{m_s(\mu) + m_d(\mu)}\right]^2 F_\pi \,B_8^{(1/2)}\,,
\end{eqnarray}
with \cite{Buras:1985yx,Bardeen:1986vp,Buras:1987wc}
\be\label{LN}
\bsi=\bei=B_8^{(1/2)}=1, \qquad {\rm (large~N~Limit)}\,.
\ee
Note, that using the definition of $B_i$ parameters consistent with the large $N$ limit of QCD, as given above, implies that their values in the VIA \cite{Buras:1993dy} read
\be\label{VIA}
\bsi=1, \qquad \bei\approx 0.99, \qquad B_8^{(1/2)}\approx 1.2 \qquad  {\rm (VIA)}\,.
\ee
We will return to this point in the next section.

The input values of parameters entering these expressions are given by
\cite{Agashe:2014kda,Aoki:2013ldr}
\be\label{FpFK}
F_\pi=130.41(20)\mev,\qquad  \frac{F_K}{F_\pi}=1.194(5)\,
\ee
\be
m_s(m_c)=109.1(2.8)\mev, \qquad  m_d(m_c)=5.44(19)\mev \,.
\ee

It should be emphasized that the overall factor $h$ in these expressions
depends on the normalisation of the amplitudes $A_{0,2}$. In \cite{Buras:1993dy}
and recent papers of the RBC-UKQCD collaboration \cite{Blum:2012uk,Blum:2015ywa}
$h=\sqrt{3/2}$ is used whereas in most recent phenomenological papers
\cite{Cirigliano:2011ny,Buras:2014maa,Buras:2014sba,Buras:2015qea}, $h=1$. 
In the present paper we will keep general $h$ so that, e.g., the decay amplitude 
$K^+\to\pi^+\pi^0$ reads $(3/2h)A_2$.

Comparing the expressions (\ref{eq:Q60}) and (\ref{eq:Q80}) with the 
lattice results in \cite{Bai:2015nea}, we find {(see also \cite{Buras:2015yba})}\footnote{To this end, the values 
 $m_s=102.27\mev$ and $m_d=5.10\mev$ at $\mu=1.53\gev$ have to be used.}
\be\label{B8LAT0}
\bsi=0.57\pm 0.19,\qquad B_8^{(1/2)}=1.0 \pm 0.2\,,\qquad  {\rm (lattice~QCD)}.
\ee
On the other hand, comparing  (\ref{eq:Q82}) with the value for this matrix 
element obtained by RBC-UKQCD collaboration in \cite{Blum:2015ywa}
one extracts \cite{Buras:2015qea}
\be\label{Lbei}
B_8^{(3/2)}=0.76\pm 0.05\,, \qquad  {\rm (lattice~QCD)}.
\ee
All these results are very weakly dependent on the renormalization scale. 
The quoted values correspond to $\mu=1.53\gev$. Basically, identical results are 
obtained for $\mu=m_c$ used in {\cite{Buras:2015yba}}. However, as stated 
before (\ref{B8LAT0}), in extracting these parameters from  \cite{Bai:2015nea} it is important to use the quark masses at that scale. 

As we will demonstrate in the next section, these lattice results are 
consistent 
with the large $N$ approach. Indeed, we will show that the following pattern emerges at next-to-leading order in our dual approach:
\begin{eqnarray}
\bsi &=& 1-\left[\frac{F_\pi}{F_K-F_\pi}\right]\ord(\frac{1}{N}) < 1\,, \label{NBOUNDS}\\
\bei &=& 1-\ord(\frac{1}{N}) < 1\,, \label{NBOUNDSa}\\
B_8^{(1/2)}&=& 1 + \ord (\frac{1}{N}) >1\,.\label{NB}
\end{eqnarray}

We would like to recall that strong indication for the suppression of $\bei$  below unity in our approach 
have been found already in 1998 in  \cite{Hambye:1998sma},
while in the case of $\bsi$ no clear cut conclusions could be reached. Our 
present analysis of both $\bsi$ and $\bei$ clearly indicates the negative signs of $1/N$ corrections 
to the leading result in (\ref{LN}).

Finally, the lattice results in \cite{Bai:2015nea} and \cite{Blum:2015ywa} 
exhibit colour suppression of the matrix elements of $Q_3$, $Q_5$ and $Q_7$
operators relative to the ones of $Q_4$, $Q_6$ and $Q_8$, respectively:
\be\label{C1}
\frac{\langle Q_3 \rangle_0}{\langle Q_4 \rangle_0}= -0.18\pm 0.25, \qquad
\left(\frac{F_K}{F_\pi}-1\right)\frac{\langle Q_5 \rangle_0}{\langle Q_6 \rangle_0}= 0.10\pm 0.05\,,
\ee
\be\label{C2}
\frac{\langle Q_7 \rangle_0}{\langle Q_8 \rangle_0}= 0.13\pm 0.04 \qquad 
\frac{\langle Q_7 \rangle_2}{\langle Q_8 \rangle_2}= 0.22\pm 0.01\,. 
\ee
These results are consistent with the large $N$ approach.
Indeed, as we will demonstrate soon, the ratios in (\ref{C1}) are $\ord(1/N^2)$
while the ratios in (\ref{C2}) are  $\ord(1/N)$.

These results  allow to simplify some of the relations 
between the matrix elements so that it is justified to use the relations
\begin{eqnarray}
\langle Q_4 \rangle_0 &=&  \langle Q_2 \rangle_0
                          -\langle Q_1 \rangle_0 \,, \label{Q4a}\\
\langle Q_9 \rangle_0 &=&
\frac{3}{2} \langle Q_1 \rangle_0  \, ,
\label{eq:5.12a} \\
\langle Q_{10} \rangle_0 &=&
    \langle Q_2 \rangle_0 + \frac{1}{2} \langle Q_1 \rangle_0,
\label{eq:5.13a}
\end{eqnarray}
which simplify  the phenomenological analysis of $\epe$ in \cite{Buras:2015yba}.

\section{Derivations}\label{sec:3}
The large $N$ numerical values of the $|\Delta S|=1$ matrix elements already 
displayed in Section~\ref{sec:3.2} are most easily derived from the effective theory for 
the the pseudo-Goldstone field
\be
U(\pi)\equiv \exp(i\sqrt{2}\frac{\pi}{f})
\ee
with $\pi=\lambda_a\pi^a$, the meson nonet lying below the one GeV and $f$, the associated weak decay constant scaling like $\sqrt{N}$.
In particular, the electroweak penguin operator introduced in (\ref{O4}) and ``Fierzed''
into a product of two colour-singlet quark densities, namely,
\be
Q_8= -12 \sum_{q=u,d,s}(\bar s_Lq_R)e_q (\bar q_R d_L)
\ee
can be hadronized by considering the leading chiral effective Lagrangian in the large $N$ limit:

\be\label{eq:44}
L_{\rm eff}(p^2,N)=\frac{f^2}{8} Tr\left[\partial_\mu U \partial^\mu U^+ +
r(m U^\dagger+U m^\dagger)\right].
\ee
Indeed, a straightforward identification of the second term in this equation with the standard Dirac mass term in QCD
\be
L_{\rm QCD}({\rm mass})=-(\bar q_L m q_R+\bar q_R m^\dagger q_L)
\ee
 allows us to hadronize all colour-singlet quark densities
\be
\bar q^a_R q_L^b= -\frac{f^2}{8} r U^{ba}
\ee

\be
\bar q^a_L q_R^b= -\frac{f^2}{8} r U^{\dagger ba}
\ee
such that
\be
Q_8=-\frac{3}{16} f^4 r^2\sum_{q=u,d,s} U^{dq} e_q  U^{\dagger qs}\,.
\ee
Consequently, the factorized matrix elements of the $\Delta I=1/2$ and $\Delta I=3/2$ components of $Q_8$ in the large $N$ limit are
\be\label{eq:49a}
\langle Q_8\rangle_0=\frac{h}{2} f r^2
\ee

\be\label{eq:50a}
\langle Q_8\rangle_2=
\frac{h}{2\sqrt{2}} f r^2 \,.
\ee

Similarly, the QCD penguin operator $Q_6$ introduced in (\ref{O3}) and ``Fierzed'' into a product of two colour-singlet densities reads
\be\label{eq:50}
Q_6= -8 \sum_{q=u,d,s}(\bar s_Lq_R) (\bar q_R d_L)= -\frac{1}{8} f^4 r^2\sum_q U^{dq}  U^{\dagger qs}\,=0\,.
\ee
As a matter of fact, one has the relation
\be\label{eq:51}
r(\mu)=\frac{2 m_K^2}{m_s(\mu)+m_d(\mu)}
\ee
at the level of $L_{\rm eff}(p^2,N)$. Yet, at this level, the absence of $SU(3)$ splitting among the weak decay
constants implies ill-defined $\langle Q_8\rangle_{0,2}$ matrix elements 
in (\ref{eq:49a}) and (\ref{eq:50a})
as 
well as a vanishing  $Q_6$ operator in (\ref{eq:50}).

It is well known \cite{Chivukula:1986du,Bardeen:1986vp}   that the next-to-leading term in the chiral effective Lagrangian
\be\label{eq:52}
L_{\rm eff}(p^4,N)=-\frac{f^2}{8} \frac{r}{\Lambda_\chi^2}Tr\left[m \partial^2 U^\dagger+ \partial^2 U m^\dagger\right]
\ee
solves both problems since it leads to realistic weak decay constants
\be
F_\pi=(1+\frac{m_\pi^2}{\Lambda_\chi^2})f
\ee
\be
\frac{F_K}{F_\pi}=1+\frac{m_K^2-m_\pi^2}{\Lambda_\chi^2}\,
\ee
thanks to its derivative dependence and, simultaneously, it implies 
\be\label{JG}
Q_6=-\frac{f^4}{4}\left(\frac{r}{\Lambda_\chi}\right)^2(\partial_\mu U\partial^\mu U^\dagger)^{ds}+\ord(\frac{1}{\Lambda_\chi^4})\approx -\left(\frac{r}{\Lambda_\chi}\right)^2 Q_4
\ee
from the shift induced in the 
hadronized quark densities 
\be
U\rightarrow U-\frac{1}{\Lambda^2_\chi}\partial^2U
\ee
through its mass dependence. Taking these corrections into account, we now reproduce the large $N$ matrix elements given in (\ref{eq:Q60})-(\ref{eq:Q80}), with the normalization (\ref{LN}) for the $B_{6,8}$  coefficients if contributions  $\ord(m_\pi^2/\Lambda_\chi^2)$ to $Q_8$ are neglected.

At this point, it is worth emphasizing that, here, we consistently normalize the
$|\Delta S|=1$ $B_{6,8}$ to unity in the large $N$ limit. Such is unfortunately not the case for the $|\Delta S|=2$ $B_K$ parameter conventionally normalized to one with respect to VIA in (\ref{eq:5}). Had the $\Delta S=2$ matrix element in
(\ref{eq:5}) been normalized relative to its large $N$ value, the most precise 
$\hat B_K$ parameter extracted from lattice QCD  in (\ref{L2012})
would read
$\hat B_K=1.021\pm 0.013$ nowadays and our result in (\ref{BKfinal}) 
$\hat B_K=0.97\pm 0.03$.
 In \cite{Buras:1993dy,Hambye:1998sma}, $B_6$ and $B_8$ were also normalized with respect to the VIA as in (\ref{VIA}).

We are now in an ideal position to estimate $1/N$ corrections encoded in the 
$B_{6,8}$ parameters. The factorizable $1/N$ corrections to $|\Delta S|=1$ density-density operators are fully included \cite{Buras:1987wc} in the running of quark masses in (\ref{eq:51}). Let us thus focus on non-factorizable one loop corrections induced by $L_{\rm eff}(p^2,N)$. Applying the background field method  of \cite{Fatelo:1994qh}, we find
\be\label{eq:55}
U^{dq} U^{\dagger q^\prime s}(\Lambda)= U^{dq} U^{\dagger q^\prime s}(M)- \frac{16}{f^4}
\frac{\ln(\Lambda^2/M^2)}{(4\pi f)^2}\left[2 J_L^{ds} J_R^{q^\prime q}+
(J_L J_L)^{ds}\delta^{q^\prime q}\right](M)
\ee
with $\Lambda=\ord(1\gev)$ the euclidean ultraviolet cut-off of the effective theory (\ref{eq:44}) to be matched with the non-factorizable short distance evolution, $M=\ord(m_K)$ and 
\be
J_L^{ab}=\bar q_L^b\gamma_\mu q_L^a=i \frac{f^2}{4} (\partial_\mu U U^\dagger)^{ab}
\ee
\be
J_R^{ab}=\bar q_R^b\gamma_\mu q_R^a=i \frac{f^2}{4} (\partial_\mu U^\dagger U)^{ab}
\ee
the colour-singlet left-handed and right-handed hadronic currents derived from  $L_{\rm eff}(p^2,N)$, respectively.

Applied to the specific $K\to\pi\pi$ decay processes,
\begin{itemize}
\item
the first (L-R) current-current operator in (\ref{eq:55}), also present with the right relative sign in the VIA through a Fierz transformation, does not contribute to the matrix element
$\langle Q_{6}\rangle_0$ since
\be
\text{Tr} (J_R)= \frac{\sqrt{3}}{2} f \partial_\mu \eta^0
\ee
\item
the second (L-L)  current-current operator in (\ref{eq:55}), absent in the VIA, does not contribute to the matrix elements $\langle Q_{8}\rangle_{0,2}$ since
\be
\text{Tr} (e_q)=0\,.
\ee
\end{itemize}

An explicit calculation of the surviving $Q_4$ and $Q_7$ matrix elements (in the large $N$ limit) gives then, respectively,
\begin{eqnarray}
\bsi &=& 1 -\frac{3}{2} \left[\frac{F_\pi}{F_K-F_\pi}\right] \frac{(m_K^2-m_\pi^2)}{(4\pi F_\pi)^2}\ln(1+\frac{\Lambda^2}{\tilde m_6^2})=1-0.66\,\ln(1+\frac{\Lambda^2}{\tilde m_6^2})\label{59a}\\
B_8^{(1/2)} &=& 1+ \frac{(m_K^2- m_\pi^2)}{(4\pi F_\pi)^2}\ln(1+\frac{\Lambda^2}{\tilde m_8^2})=1+0.08\, \ln(1+\frac{\Lambda^2}{\tilde m_8^2})
\label{59b}\\
\bei &=& 1 -{2} \frac{(m_K^2- m_\pi^2)}{(4\pi F_\pi)^2}\ln(1+\frac{\Lambda^2}{\tilde m_8^2})=1-0.17\, \ln(1+\frac{\Lambda^2}{\tilde m_8^2})\label{59c}
\end{eqnarray}
with pseudoscalar mass scale parameters bounded necessarily by the effective cut-off around $1\gev$:
\be
\tilde m_{6,8}\le \Lambda\,.
\ee

First, we emphasize most important properties of these results:
\begin{itemize}
\item
For $\Lambda=0$, corresponding to strict large $N$ limit and matrix elements 
evaluated at zero momentum, $\bsi=\bei=B_8^{(1/2)}=1$ in accordance with (\ref{LN}). 
\item
With increasing $\Lambda$, the parameters $\bsi$ and $\bei$ decrease below unity  and $\bsi$ decreases faster than $\bei$. Consequently, at scales $\ord(1\gev)$ 
relevant for the phenomenology both $\bsi$ and $\bei$ are predicted to be 
below unity and there is strong indication that $\bsi < \bei$. 
\item
While the dependence of  $\bsi$ and $\bei$ on $\Lambda < 1\gev$ is stronger 
than their dependence on $\mu$ in the perturbative regime, these two
properties of $\bsi$ and $\bei$ are at the qualitative level consistent 
with the numerical analysis performed for  $\bsi$ and $\bei$ 
by means of the standard renormalization group running in \cite{Buras:1993dy}. 
Indeed as seen in Figs. 11 and 12 of that paper $\bsi$ decreases  with increasing $\mu$, faster than $\bei$, albeit in this perturbative range the dependence of $\bsi$ and $\bei$ on $\mu$ is very weak. While the analysis in  \cite{Buras:1993dy} includes NLO QCD and QED corrections, the inspection of the one-loop anomalous dimension matrix allows to see these properties explicitly. In particular 
$Q_6$ mixes with the linear combination $(Q_4+Q_6)$ and
we find for $\mu_1\le \mu_2\le m_c$
\be
\bsi(\mu_2)= \bsi(\mu_1)\left[1-\frac{\alpha_s(\mu_1)}{2\pi}\ln(\frac{\mu_2}{\mu_1})\left(1+\frac{\langle Q_4(\mu_1)\rangle_0}{\langle Q_6(\mu_1)\rangle_0}\right)\right]\,.
\ee
From (\ref{JG}) $|\langle Q_6(\mu_1)\rangle_0|>|\langle Q_4(\mu_1)\rangle_0|$ such that  $\bsi$ decreases with increasing $\mu$. On the other hand, in the LO 
$Q_8$ runs only by itself and the one-loop anomalous 
dimension matrix implies 
\be
B_8^{(1/2,3/2)}(\mu_2)=B_8^{(1/2,3/2)}(\mu_1)\,,
\ee
which follows from exact $SU(3)$ symmetry imposed in SD calculations. The breakdown of $SU(3)$ is only 
 felt in the matrix elements of $Q_8$ making in the LD range $\bei$ dependent 
weakly on the scales involved. 
In view of this, the suppression of both $\bsi$ and $\bei$ below the unity 
can be considered as a solid result and our explicit calculation as well 
as different behaviour of $Q_6$ and $Q_8$ under flavour $SU(3)$ provide
 a strong 
support for $\bsi < \bei$.  On the other hand, 
\be
B_8^{(1/2)}\approx [\bei]^{-1/2}
\ee
 slightly increases with $\Lambda$ which is also consistent with 
the standard renormalization group running \cite{Buras:2015yba}.
\end{itemize}

Next, we observe that:
\begin{itemize}
\item
The numerical value of the parameter $\bsi$ suffers from rather large uncertainties. This feature is related to the fact 
that $Q_6$ vanishes at leading order in chiral perturbation theory (see (\ref{eq:50})).
The $1/N$ logarithmic correction in (\ref{59a}) is therefore 
{\it artificially enhanced} by the factor $F_\pi/(F_K-F_\pi)\approx 5$ such that
\be
\bsi < 0.6 \,.
\ee
\item
The parameter $B_8^{(1/2)}$ has a very small $1/N$ correction. At $\ord(1/N^2)$, one larger
 contribution might arise from the anomalous effective Lagrangian
\be
L_{\rm eff}(p^0,1/N)= \frac{f^2}{32}\left(\frac{m_0^2}{N}\right)\left[Tr(\ln U-\ln U^\dagger)\right]^2
\ee
that solves the so-called $U(1)_A$ problem \cite{Weinberg:1975ui} by providing the $\eta^\prime$ 
pseudoscalar with a physical  mass in the large $N$ limit \cite{Gerard:2004gx}:
\be
m_{\eta^\prime}^2+m_{\eta}^2- 2 m_K^2\approx m_0^2\approx 0.7 \gev^2\,.
\ee

Applying again the background field method, we obtain
\be
U^{dq}U^{\dagger q^\prime s}(\Lambda)=\left[1-\frac{4}{N}\frac{m_0^2}{(4\pi f)^2} \ln (\frac{\Lambda^2}{M^2})\right]U^{dq}U^{\dagger q^\prime s}(M).
\ee
Such a negative contribution to $Q_8$ has been included in \cite{Hambye:1998sma}.
However, any consistent estimate beyond 
\be
B_8^{(1/2)}\approx 1
\ee
would require a full calculation at $\ord(1/N^2)$.
\item
The parameter $\bei$, for which the $1/N$ expansion is more reliable,  is found  in the range 
\be\label{rangebsi}
0.7 \le \bei \le 0.9 \,
\ee
\end{itemize}
if $\tilde m_8\ge m_K$.

Lattice result for $\bsi$ in (\ref{B8LAT0}) turns out to almost saturate our 
bound. But one should realize that although we are confident about 
the suppression of $\bsi$ below unity, its actual size is rather uncertain. 
For instance the inclusion of dynamical scalars presently frozen in 
$\Lambda_\chi$ could reduce the coefficient in front of the logarithm in (\ref{59a}) making $\bsi$ larger. This uncertainty in the value of $\bsi$ explains also
why it took so long  to calculate $\bsi$ in lattice QCD even with a
large uncertainty as seen in  (\ref{B8LAT0}). On the other hand, the range for $\bei$ in (\ref{rangebsi}) is consistent with 
the one in (\ref{Lbei}).  These results indicate that indeed $\bsi$ could be 
smaller than $\bei$. Yet, in view of the large numerical uncertainties in 
the case of $\bsi$, we cannot exclude that $\bsi$ is as large as $\bei$. 
We therefore believe that the best way of summarizing our results for 
$\bsi$ and $\bei$ is given in (\ref{NBOUNDS}) and (\ref{NBOUNDSa}) together 
with 
\be
\bsi \le \bei < 1\,.
\ee

Below $1\gev$ we have seen in (\ref{eq:55}) that density-density operators transmute into current-current ones at 
$\ord(1/N)$. But power counting in our effective theory does not allow the other  way around, namely current-current 
operators evolving into density-density ones. This is fully consistent 
with the evolution of hadronic matrix elements above $\mu=1\gev$ studied 
already in \cite{Buras:1993dy} 
 and is opposite to the evolution of the corresponding 
Wilson coefficients. Now, in the large $N$ approach, it has already been 
shown \cite{Buras:2014maa,Fatelo:1994qh} that any (L-L) current-current operator  evolves as 
\be\label{G1}
J_L^{ab} J_L^{cd}(\Lambda)=
J_L^{ab} J_L^{cd}(0)- \ord(\frac{1}{N})\left[ 2 J_L^{ad} J_L^{cb}-\delta^{ad}(J_L J_L)^{cb} -\delta^{cb}(J_L J_L)^{ad}\right](0)\,
\ee
to stand in contrast with the wrong relative sign in the VIA analogue
\be\label{VIAG1}
J_L^{ab} J_L^{cd}(\Lambda)=
J_L^{ab} J_L^{cd}(0)+ \frac{1}{N} J_L^{ad} J_L^{cb}(0) \, .
\ee
As a consequence, summing over $c=d=u,d,s$ we conclude that the matrix element $\langle Q_3\rangle_0$ which vanishes in the large $N$ limit is formally 
$\ord(1/N^2)$ relative to $\langle Q_4\rangle_0$ in our effective theory. Such a strong suppression could have been anticipated from the LO short-distance evolution of the four-quark operator $Q_3$ into another linear combination of $Q_4$ and $Q_6$:
\be
Q_3(\mu_2)= Q_3(\mu_1) - \ord(\frac{1}{N})\left[\frac{11}{2} Q_4+ Q_6\right](\mu_1).
\ee
At long distance, the further $Q_3$ evolution undergoes an important numerical cancellation since (\ref{JG}) tells us that the $Q_4$ and $Q_6$ operators are not independent anymore. Following (\ref{G1}), this numerical cancellation is not a mere coincidence but the result of a consistent $1/N$ expansion.

In the same manner, it has been proved \cite{Fatelo:1994qh} that any (L-R) 
current-current operator evolves as 
\be\label{DD}
J_L^{ab} J_R^{cd}(\Lambda)=
J_L^{ab} J_R^{cd}(0) +
\ord(\frac{1}{N})\left[U^{ad} (\delta U^\dagger)^{cb}+\delta U^{ad}(U^\dagger)^{cb} \right](0)\,
\ee
with $\delta U$ proportional to $(\Box U-U\Box U^\dagger U)$.

Consequently, the matrix element $\langle Q_{5}\rangle_0$  which  vanishes in the large $N$ limit is $\ord(1/N^2)$ 
  relative to 
$F_\pi/(F_K-F_\pi) \langle Q_6\rangle_0$ in our effective theory.
As already mentioned, an enhancement factor has to be introduced to compensate for the ``accidental'' chiral suppression of  $\langle Q_6\rangle_0$ in (\ref{eq:50}). On the other hand, the matrix elements
\be\label{xx}
\langle Q_{7}\rangle_{0}=\frac{h}{2} F_\pi(m_K^2-m_\pi^2)
\ee
\be\label{yy}
\langle Q_{7}\rangle_{2}=-\frac{h}{\sqrt{2}} F_\pi(m_K^2-m_\pi^2)
\ee
are  $\ord(p^2)$ but non zero in the large $N$ limit. With the matrix elements $\langle Q_8\rangle_{0,2}$  given in (\ref{eq:49a}), (\ref{eq:50a}) and   being $\ord(p^0)$, the 
resulting ratios $\langle Q_7\rangle_{0,2}/\langle Q_8\rangle_{0,2}$ are 
at the level of a few percent  and can thus  be neglected. The long-distance 
evolution of $Q_7$ in (\ref{DD}) leads then to matrix elements proportional to 
$\langle Q_8 \rangle_{0,2}$, though $\ord(p^2)$. This is clearly 
at variance with 
 its LO short-distance evolution, namely 
\be
Q_7(\mu_2)=
Q_7(\mu_1)+ \ord(\frac{1}{N})\, Q_8(\mu_1)\,.
\ee
As already explicitly stated in  \cite{Fatelo:1994qh}, this suggests the necessity to introduce higher resonances beyond our effective theory truncated to the low-lying pseudoscalars. In a dual representation of QCD the matrix elements  $\langle Q_7\rangle_{0,2}$ should then be dominantly  $\ord(p^0)$, but $1/N$-suppressed, with the bound
\be
\frac{\langle Q_7\rangle_0}{\langle Q_7\rangle_2}<\sqrt{2}
\ee
resulting from the isospin decompositions in (\ref{xx})-(\ref{yy}) and 
(\ref{eq:49a})-(\ref{eq:50a}).

\boldmath
\section{Implications for $\epe$}\label{sec:5}
\unboldmath
We will now briefly discuss the implications of our results for $\epe$. To this 
end we will use the analytic formula for $\epe$ in the SM derived recently in
\cite{Buras:2015yba}. In obtaining this formula it has been assumed that the SM describes exactly the data on CP-conserving
$K\to \pi\pi$ amplitudes: ${\rm Re} A_0$ and ${\rm Re} A_2$. This allowed to
determine  the contributions of the $(V-A)\otimes (V-A)$ QCD penguin operator
$Q_4$ and of the electroweak penguin operators $Q_9$ and $Q_{10}$ to $\epe$
much more precisely than it is presently possible by lattice QCD and large $N$ 
approach. This determination was facilitated by our results on the suppression 
of the matrix element $\langle Q_3 \rangle_0$ implying the relations 
(\ref{Q4a})-(\ref{eq:5.13a})\footnote{The final numerical analysis in  \cite{Buras:2015yba} leading to 
(\ref{AN2015}) included also small corrections from $Q_3$ and other corrections from subleading operators.}.

The formula in question reads \cite{Buras:2015yba}
\begin{equation}
\frac{\varepsilon'}{\varepsilon} =  10^{-4} \biggl[
\frac{\IM\lambda_{\rm t}}{1.4\cdot 10^{-4}}\biggr] \left[\,
a\big(1-\hat\Omega_{\rm eff}\big) \big(-4.1 + 24.7\,\bsi\big) + 1.2 -
10.4\,\bei \,\right] ,
\label{AN2015}
\end{equation}
where
\be
\IM\lambda_{\rm t}=\IM ({V_{td}V_{ts}^*})= \vub\vcb \sin\gamma
\ee
and \cite{Cirigliano:2003nn,Cirigliano:2003gt,Bijnens:2004ai,Buras:2015yba}
\be
a=1.017, \qquad \hat\Omega_{\rm eff} = (14.8 \pm8.0)\times 10^{-2}\,.
\ee
$\vub$ and $\vcb$ are the elements of the CKM matrix and $\gamma$ is an
angle in the unitarity triangle.
The parameters $a$  and $\hat\Omega_{\rm eff}$  represent isospin breaking corrections \cite{Cirigliano:2003nn,Cirigliano:2003gt,Bijnens:2004ai}.  See these papers 
 and \cite{Buras:2015yba} for details.

Setting all parameters, except for $\bsi$ and $\bei$, in (\ref{AN2015}) to their
 central values 
we find
\begin{eqnarray}
\RE(\epe)&=& 8.6 \times 10^{-4},\qquad  (\bsi=1.0, \bei=1.0)\,\\
\RE(\epe)&=& 6.4 \times 10^{-4},\qquad  (\bsi=0.8, \bei=0.8)\,\\
\RE(\epe)&=& 2.2 \times 10^{-4},\qquad  (\bsi=0.6, \bei=0.8)\,.
\end{eqnarray}
A detailed anatomy of $\epe$ in the SM is presented in \cite{Buras:2015yba}, where 
various uncertainties related to  NNLO QCD corrections and other  uncertainties 
are discussed. But these three examples indicate that taking our bounds into 
account and guided by the results on $\bsi$ and $\bei$ from lattice QCD 
and our dual approach, our SM prediction for $\epe$ appears to be significantly below the data given in (\ref{eprime}).

In each of these predictions, there are uncertainties  from 
the value of $\IM\lambda_t$, the unknown complete NNLO corrections to Wilson coefficients of contributing operators, $\alpha_s$, $m_t$ and other input parameters. But they appear not to change the conclusion that, presently, the SM prediction for $\epe$ is significantly below the data. Our upper bound on $\bsi$ plays an important role in this result as otherwise increasing $\bsi$ above unity would allow to fit easily the data. 

\boldmath
\section{Comments on other large $N$ QCD approaches}\label{otherN}
\unboldmath
In \cite{Bijnens:2000im} the authors analyse $\Delta I=1/2$ rule and  $\epe$ 
in the chiral limit including $1/N$ corrections. Their results differ drastically from our results. In particular, their low energy ''Extended Nambu-Jona-Lasinio'' model (ENJL) gives
\be
\bsi\approx 3, \qquad \bei\approx 1.3, \qquad ({\rm ENJL})
\ee
namely  $\bsi$ roughly by a factor of five larger than lattice calculations and our results.
With such high values of $\bsi$, QCD penguins play an important role in the explanation of the $\Delta I=1/2$ rule and  the experimental value of $\epe$ can easily be reproduced, again in contrast with lattice QCD and our 
dual QCD approach.

In \cite{Hambye:2003cy} the authors consider a low energy model including the 
light pseudo-scalar, vector and scalar poles only in the chiral limit. Within 
this so-called ''Minimal Hadronic Approximation'' (MHA), they also obtain much larger  values of $\bsi$ and $\bei$ than in our approach 
\be
\bsi\approx 3, \qquad \bei\approx 3.5, \qquad ({\rm MHA})
\ee
and find then good agreement with data for $\epe$.

In \cite{Bijnens:2001ps,Cirigliano:2001qw,Cirigliano:2002jy} the authors 
rely on dispersion relations and ''Finite Energy Sum Rules'' (FESR) in the chiral limit to extract the  electroweak penguin matrix elements from ALEPH and OPAL
 data. Doing so, they obtain central values for $\bei$ by  a factor of two to three larger than in lattice QCD and our dual QCD approach
\be
1.3\le\bei\le 2.5, \qquad ({\rm FESR})\,.
\ee

We conclude that the recent lattice QCD results tend to demonstrate that the 
various models considered in  
\cite{Bijnens:2000im,Hambye:2003cy,Bijnens:2001ps,Cirigliano:2001qw,Cirigliano:2002jy} do not represent properly the low energy QCD dynamics at work for 
penguin matrix elements, but confirm 
the structure of our dual QCD approach.

\section{Summary and outlook}\label{sec:4}
In the present paper, we have compared the structure of the hadronic matrix 
elements in $K\to\pi\pi$ decays obtained within the dual approach to QCD 
with the one obtained recently by the RBC-UKQCD lattice approach to QCD and commented 
briefly on the status of the parameter $\hat B_K$ and the $\Delta I=1/2$ 
rule. Our main results are as follows:
\begin{itemize}
\item
The status of $\hat B_K$ is very good as both our approach and the lattice QCD
calculations give this parameter very close to $0.75$. But we expect that 
 $\hat B_K$ from the lattice approach will decrease by a few $\%$ in the 
coming years.
\item
While the results for ${\rm Re}A_2$ obtained in both approaches agree well
with the data, the central value of ${\rm Re}A_0$ from RBC-UKQCD collaboration 
is by a factor ot two larger than in our approach and $40\%$ above the data.
While our result in (\ref{DRULEN}) appears from present perspective to be 
final in our approach, significant improvement on the lattice result is expected
in the coming years. This will allow to find out whether at some level of 
$20\%$ new physics could still be responsible for the $\Delta I=1/2$ rule. 
An analysis anticipating such possibility has been presented in  \cite{Buras:2014sba}.
\item
As the upper bound on $\bei$ in (\ref{NBOUNDSa}) has been already indicated 
in  \cite{Hambye:1998sma}, one of the most important results of our paper is 
the upper bound on $\bsi$. Our estimate suggests that $\bsi\le \bei<1$,
but the precise values can only be obtained by lattice methods. 
\item
Among other results of our approach supported by recent results from RBC-UKQCD
is the strong suppression of $\langle Q_{3,5}(\mu)\rangle $ and $B_8^{(1/2)}\approx 1$.
\end{itemize}

 If indeed the emerging pattern
$\bsi\le \bei<1 $ with $\bei=0.8\pm0.1$ will be confirmed by more precise calculations one day, the very recent analysis in \cite{Buras:2015yba} and our paper show that $\epe$ within the SM will be found roughly by  a factor of two below the data. 
For a detailed phenomenological discussion of the state of $\epe$ within the 
SM including all errors and future theoretical and experimental
prospects we refer to \cite{Buras:2015yba}. On the other hand, first phenomenological implications of our results on new physics models have been presented in 
\cite{Blanke:2015wba,Buras:2015yca}.

\section*{Acknowledgements}
We thank Martin Gorbahn, Sebastian J{\"ager}, Matthias Jamin, Tadeusz Janowski and Chris Kelly for discussions.
This research was done and financed in the context of the ERC Advanced Grant project ``FLAVOUR''(267104) and the Belgian IAP Program BELSPO P7/37. It was also
 partially
supported by the DFG cluster
of excellence ``Origin and Structure of the Universe''.

\renewcommand{\refname}{R\lowercase{eferences}}

\addcontentsline{toc}{section}{References}

\bibliographystyle{JHEP}
\bibliography{Neprefs}
\end{document}